\documentclass[10pt,a4paper]{article}

\usepackage{authblk}
\usepackage{graphicx} 
\usepackage{graphics}
\usepackage{caption}
\usepackage{subcaption}
\usepackage{float}
\usepackage[section]{placeins}
\usepackage{lastpage}
\usepackage[format=plain,justification=raggedright,singlelinecheck=false,font=small,labelfont=bf,labelsep=space]{caption} 
\usepackage{fancyhdr}
\usepackage{color}
\usepackage{latexsym}
\usepackage{amsmath}
\usepackage{amssymb}
\usepackage{amsfonts}
\usepackage{bm}

\begin{document}



\title{Wake and wave resistance on viscous thin films}

\author[1,3]{Ren\'{e} Ledesma-Alonso}
\author[1,2]{Michael Benzaquen}
\author[1]{Thomas Salez}
\author[1]{Elie Rapha\"{e}l}

\affil[1]{Laboratoire de Physico-Chimie Th\'{e}orique, UMR CNRS 7083 Gulliver, ESPCI ParisTech, PSL Research University, 10 Rue Vauquelin, 75005 Paris, France.}
\affil[2]{Capital Fund Management, 23 rue de l'Universit\'{e}, 75007 Paris, France.}
\affil[2]{\texttt{rene.ledesma-alonso@espci.fr}}
\maketitle

\begin{abstract}
The effect of an external pressure disturbance, being displaced with a constant speed along the free surface of a viscous thin film, is studied theoretically in the lubrication approximation in one- and two-dimensional geometries. In the comoving frame, the imposed pressure field creates a stationary deformation of the interface - a wake - that spatially vanishes in the far region. The shape of the wake and the way it vanishes depend on both the speed and size of the external source and the properties of the film. The wave resistance, namely the force that has to be externally furnished in order to maintain the wake, is analyzed in detail. For finite-size pressure disturbances, it increases with the speed, up to a certain transition value above which a monotonic decrease occurs. The role of the horizontal extent of the pressure field is studied as well, revealing that for a smaller disturbance the latter transition occurs at a higher speed. Eventually, for a Dirac pressure source, the wave resistance either saturates for a 1D geometry, or diverges for a 2D geometry.
\end{abstract}

\section{Introduction}
\label{Sec:Intro}

A disturbance moving atop a fluid or a soft medium deforms the interface shape of the latter, thereby producing a wake.
This is a common phenomenon, which takes place in various natural, scientific and industrial settings~\cite{Carusotto2013} across several orders of magnitude, ranging from waves generated at the surface of water by animals and watercrafts~\cite{Kelvin1887}, to patterns observed during the deposition of thin polymer coatings~\cite{KistlerSchweizer}.
For instance, at the geophysical scale, the study of gravity-driven flows around human-made obstacles and landscape singularities is essential to understand and minimize the damages caused by natural disasters, such as snow avalanches~\cite{Glenne1987}, landslides~\cite{Campbell1989} and lava or mud flows~\cite{Huppert1980,Huppert1982,Kondic2003}.
In nanophysics, gaining insight on the intrusive effect of local probes, such as atomic force microscopy~\cite{Ledesma2013,Ledesma2014,Wedolowski2015}, is crucial in order to improve further the measurement accuracy. 

Similarly, the wake angle and the wave pattern generated behind a disturbance sailing at the surface of an inviscid liquid remain current topics of interest~\cite{Rabaud2013,Darmon2014,Benzaquen2014}, because they are crucial for naval industry.
In fact, to be maintained, the wake continually consumes energy, the latter being radiated away from the disturbance.
This loss can be formulated into a force, called the wave resistance~\cite{Havelock1918}, which needs to be furnished by the controller in order to ensure constant speed.
The wave resistance associated to the capillary-gravity wakes has thus been fully characterized for an incompressible inviscid flow~\cite{Raphael1996,Burghelea2002,Benzaquen2011}, including the cases of Dirac and finite-size pressure distributions.
The perturbative role of viscosity on the inertial wave resistance~\cite{Raphael1999} has also been considered, as well as the effect of thickness for a viscous liquid film~\cite{Wedolowski2013}.

In the context of viscous thin films~\cite{Oron1997,Craster2009,Blossey}, a moving external disturbance and the associated wake at the free surface may directly be used as a new kind of fine rheological probe~\cite{Decre2003,Alleborn2004Appl}.
This idea was also proposed as a possible method to measure slip at the solid-liquid boundary~\cite{Alleborn2007}.
Interestingly, despite the broad applicability of the lubrication theory and its importance for industrial processes, the study of the wake and the wave resistance in such a context is a topic which has been hastily explored and remains an open question.
For instance, the competition between visco-capillary relaxation and perturbation speed, which might control the extent and lifespan of the wake generated by the moving perturbation, has not been studied.
Imagine that the footprint left behind by the moving perturbation is arrested by performing a local quenching, before the relaxation of the thin-film occurs~\cite{Orchard1961,McGraw2012}: this may allow to imprint a narrow channel in the film, thus revealing the foundations of a new nano-patterning technique.
Additionally, one may be interested in estimating and controlling the amount of energy that this technique requires, for which the wave resistance of a moving disturbance atop a viscous thin film must be well characterized in detail.

In the present study, the surface pattern generated by a moving external pressure disturbance along a viscous thin film is described theoretically, within the lubrication approximation, for one- and two-dimensional systems. The wave resistance is analyzed in both cases, leading to analytical expressions of this force in terms of the disturbance speed and size, and the physical properties of the liquid film. Low- and high-speed regimes are described through asymptotic expressions, which unveil the nature of the wave resistance in the lubrication context.

\section{General case}
\subsection{Thin-film equation for a moving pressure disturbance}
\label{Sec:TFE}

\begin{figure}
\centering
\includegraphics[width=1\textwidth]{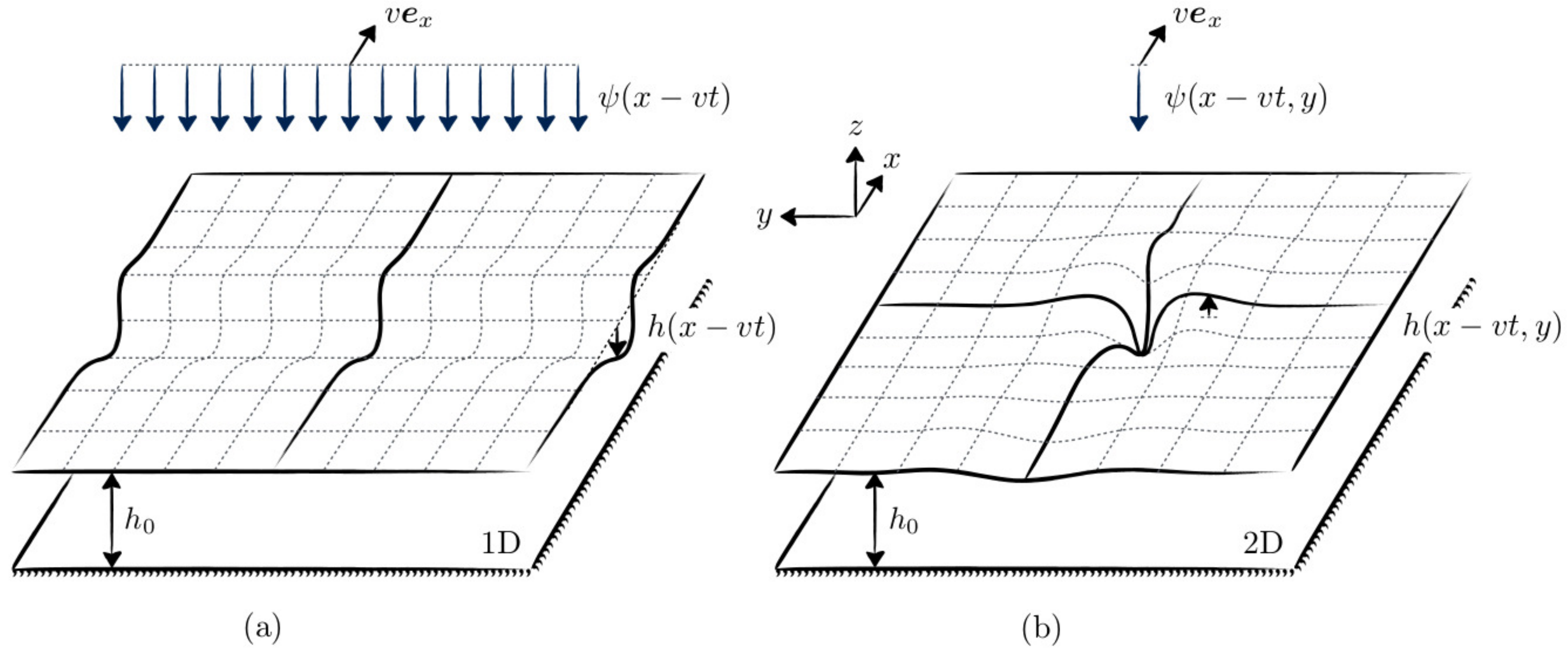}
\caption{Schematic diagram of the surface profile $h$ of a viscous thin film reacting to an external pressure disturbance $\psi$ moving along time $t$ at constant speed $v$. (a) 1D and (b) 2D geometries.}
\label{Fig:Diagram}
\end{figure}

We consider a viscous liquid film of thickness $h_0$ deposited over a flat horizontal substrate.
The equilibrium free surface of the film is flat as well, and perpendicular to the vertical direction $z$.
An external pressure field $\psi$, moving along the horizontal direction $x$  with constant speed $v\geq 0$, is applied over the liquid film.
As a result, the incompressible Newtonian fluid flow yields a stationary surface pattern $h$ (see Fig.~\ref{Fig:Diagram}) in the frame of reference of the moving disturbance.

In the lubrication approximation, neglecting shear stress at the liquid-gas interface and assuming no slip at the substrate, the incompressible Stokes equation and conservation of volume lead to the following Reynolds equation~\cite{Oron1997}:
\begin{equation}
\dfrac{\partial h}{\partial t}=\bm{\nabla} \cdot \left[ \dfrac{\left(h_0+h \right) ^3}{3\mu}\bm{\nabla} p_{\textrm{lg}} \right]\ ,
\label{TFE:Lub}
\end{equation}
where $\mu$ is the dynamic viscosity of the liquid and $p_{\textrm{lg}}$ is the pressure jump at the liquid-gas interface.
Under the lubrication hypothesis for the no-slip case, one can also neglect the normal viscous stress at the liquid-gas interface, and the pressure jump is thus given by the Young-Laplace equation.
For small slopes, the latter reads:
\begin{equation}
p_{\textrm{lg}}= -\gamma \Delta h +\rho g h+\psi\ ,
\label{TFE:Young}
\end{equation}
where $\Delta$ denotes the Laplacian operator in cartesian coordinates, $g$ is the acceleration of gravity, and $\gamma$ and $\rho$ are the liquid-gas surface tension and density difference, respectively.
Equations  \eqref{TFE:Lub} and \eqref{TFE:Young} are coupled, and should be solved together in order to describe the dynamic response of the viscous thin film due to the displacement of the external pressure field.

We introduce the dimensionless variables:
\begin{align}
X &=\kappa x \ , & Y &=\frac{y}{\ell} \ , & H &=\frac{h}{h_0} \ , \notag \\
T &=\frac{t}{\tau} \ , &  \Psi &= \frac{\psi \ell}{s_{\textrm{ext}} \kappa} \ , & S_{\textrm{ext}} &=\frac{s_{\textrm{ext}}\kappa}{\rho g h_0\ell} \ ,
\label{TFE:Adim}
\end{align}
where the capillary length $\kappa^{-1} =\sqrt{\gamma/(\rho g)}$ has been chosen as the characteristic length scale in the $x$-direction, and $\ell$ is a reference length scale in the $y$-direction, which depends on the dimension of the system: in 1D one sets $\ell=L_y$, a unit length, and in 2D one sets $\ell=\kappa^{-1}$.
In turn, $\tau =3\mu\gamma /\left[(\rho g)^2 h_0^3\right]$ is the visco-capillary time scale, and $s_{\textrm{ext}}=\int\int dxdy\,\psi$ is the imposed load, where in 1D the integral over $y$ is restricted to the horizontal extent $L_y$.
In the limit of small surface deformations $H\ll 1$, the thin-film equation (TFE), obtained from the combination of eqs.~\eqref{TFE:Young} and \eqref{TFE:Lub}, is given by:
\begin{equation}
\dfrac{\partial H}{\partial T}=-\bm{\nabla}\cdot \left[\bm{\nabla}\left(\Delta H\right)-\bm{\nabla} H-S_{\textrm{ext}}\bm{\nabla}\Psi\right]\ .
\label{TFE:Time}
\end{equation}

Let us introduce the capillary number $Ca=\mu v/\gamma$ and the Bond number $Bo=\rho g a^2/\gamma$, where $a$ denotes the characteristic horizontal size of the external pressure field.
We now place ourselves in the frame of reference of the moving disturbance, through the new variable $U=X-VT$, where the reduced speed $V$ reads:
\begin{equation}
V=v\tau\kappa\ .
\end{equation}
By assuming stationarity of the surface profile in this comoving frame, one has $H\left(X,Y,T\right)=\zeta\left(U,Y\right)$ and thus eq.~(\ref{TFE:Time}) reduces to the partial differential equation:
\begin{align}
\left[\dfrac{\partial^2}{\partial U^2}+\dfrac{\partial^2}{\partial Y^2}\right]^2\zeta-\left[\dfrac{\partial^2}{\partial U^2}+\dfrac{\partial^2}{\partial Y^2}\right]\zeta - V\dfrac{\partial \zeta}{\partial U}
=S_{\textrm{ext}}\left[\dfrac{\partial^2}{\partial U^2}+\dfrac{\partial^2}{\partial Y^2}\right]\Psi\ .
\label{TFE:2D}
\end{align}
In the following, we refer to eq.~\eqref{TFE:2D} as the 2D-TFE.

Similarly, for a 1D geometry, in which the system is invariant with respect to the $Y$-direction, eq.~\eqref{TFE:Time} is reduced to the ordinary differential equation:
\begin{equation}
\dfrac{\textrm{d}}{\textrm{d} U}\left[\dfrac{\textrm{d}^3\zeta}{\textrm{d}U^3}-\dfrac{\textrm{d} \zeta}{\textrm{d} U}-V\zeta \right]= S_{\textrm{ext}}\dfrac{\textrm{d}^2 \Psi}{\textrm{d}U^2}\ .
\end{equation}
After integrating once with respect to $U$, and considering the boundary conditions $\zeta=0$, $\textrm{d}\zeta/\textrm{d}U=0$, $\textrm{d}^3\zeta/\textrm{d}U^3=0$, and $\textrm{d} \Psi/\textrm{d}U=0$ at $\vert U\vert\rightarrow\infty$ -- since the liquid-gas interface should be undisturbed far from the moving disturbance, and since the flow and pressure gradients vanish in the far field -- one gets:
\begin{equation}
\dfrac{\textrm{d}^3\zeta}{\textrm{d}U^3}-\dfrac{\textrm{d} \zeta}{\textrm{d}U}-V\zeta=S_{\textrm{ext}}\dfrac{\textrm{d} \Psi}{\textrm{d}U}\ .
\label{TFE:1D}
\end{equation}
In the following, we refer to eq.~\eqref{TFE:1D} as the 1D-TFE.

\subsection{Wave resistance}
The wave resistance $r$ is the force that has to be externally furnished by the operator -- \textit{i.e.} the pressure field -- in order to maintain the wake. According to Havelock's formula~\cite{Havelock1918}, it reads:
\begin{equation}
r=\int\int \psi\left(x,y\right) \dfrac{\partial h}{\partial x} dx\, dy\ ,
\end{equation}
where in the 1D case the integral over $y$ is restricted to the chosen horizontal extent $L_y$. Defining the dimensionless wave resistance $R$ through:
\begin{equation}
R=\frac{r}{\rho gh_0^2\ell}\ ,
\end{equation}
and recalling eq.~\eqref{TFE:Adim}, one gets in the comoving frame:
\begin{equation}
R=\frac{S_{\textrm{ext}}}{\kappa\ell}\int\int \Psi\left(U,Y\right) \dfrac{\partial\zeta}{\partial U} dU\,dY\ .
\label{WR:Have}
\end{equation}

In the next two sections, we solve the 1D-TFE and 2D-TFE using Fourier analysis, and study the shape of their solutions and the wave resistance for different values of the two relevant dimensionless parameters: the reduced speed $V$ and the Bond number $Bo$.
One should keep in mind that $\sqrt{Bo}=a\kappa$ is the rescaled lateral extent of the external pressure field.
Also, note that, by linearity of the response, all the results correspond to an arbitrary third dimensionless number $S_{\textrm{ext}}$, the dimensionless load, as it can be absorbed in $\zeta$.
In the following, $S_{\textrm{ext}}$ is chosen to be positive, although the physical foundations and the wave resistance are valid for any sign of $S_{\textrm{ext}}$.

\section{One-dimensional case}
\label{Sec:1D}

\subsection{Surface profile}
\label{Sec:1DSurf}

\begin{figure}
\centering
\textcolor{white}{space}
\includegraphics[width=0.6\textwidth]{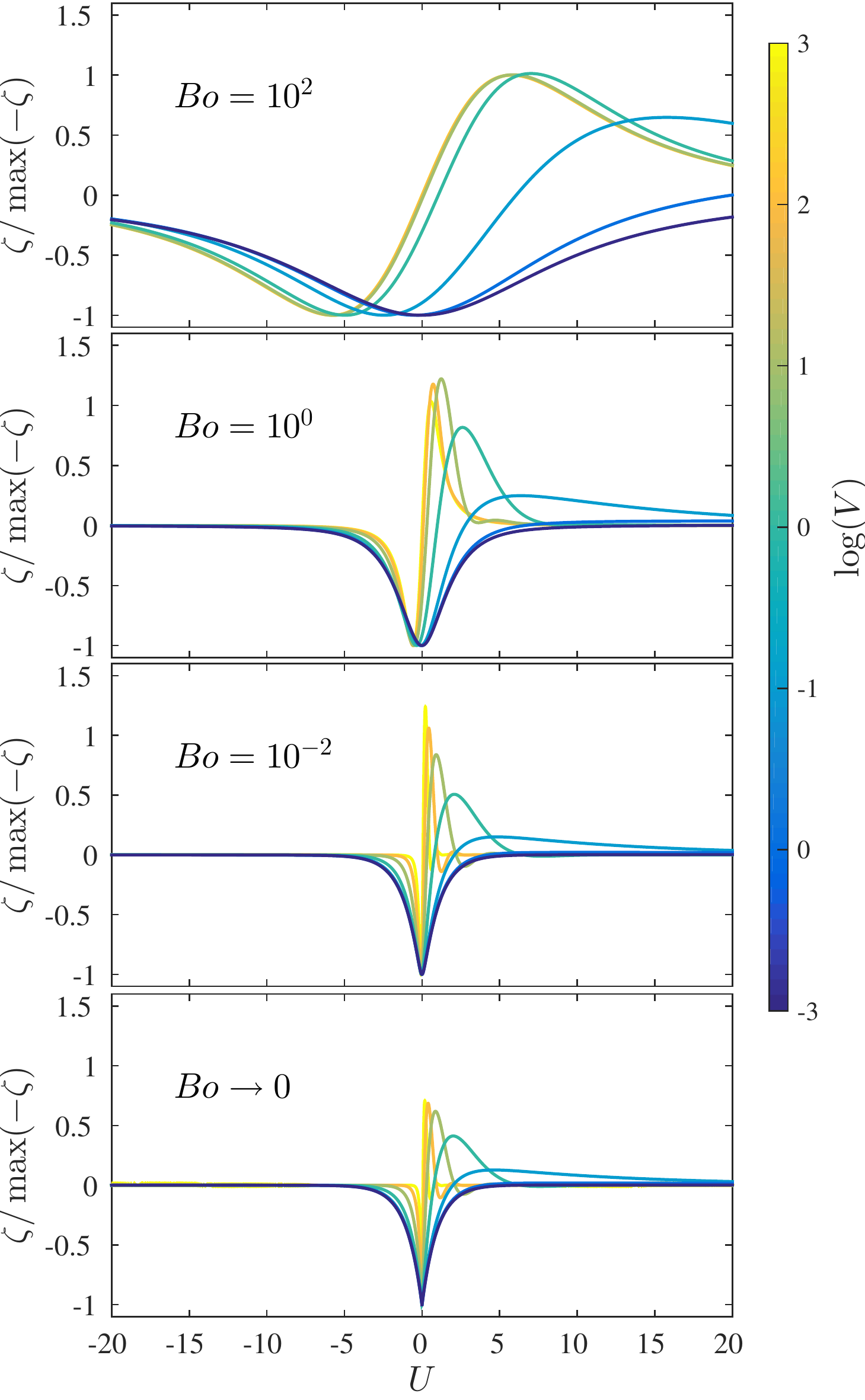}
\caption{Normalized surface profile $\zeta(U)$, solution of eq.~\eqref{TFE:1D} as given by eq.~\eqref{1D:Profile}, using Lorentzian (see eq.~\eqref{1D:Lorentz}) and Dirac ($Bo\rightarrow0$) 1D pressure distributions, with dimensionless load $S_{\textrm{ext}}>0$, for different values of the reduced speed $V$ and the Bond number $Bo$, as indicated.}
\label{Fig:1DProfiles}
\end{figure}

Considering the one-dimensional Fourier transforms $\widehat{\zeta}\left(Q\right)$ and $\widehat{\Psi}\left(Q\right)$ (see App.~\ref{App:1DFT}) of the surface profile and the pressure field, the 1D-TFE given in eq.~\eqref{TFE:1D} becomes:
\begin{equation}
-\left(iQ^3+iQ+V\right)\widehat{\zeta}=iQS_{\textrm{ext}}\widehat{\Psi}\ ,
\label{1D:TFEFT}
\end{equation}
where $i=\sqrt{-1}$ is the imaginary unit, and $Q$ is the angular wavenumber in the $U$-direction.
As a consequence, the profile $\zeta$ is described by:
\begin{equation}
\zeta\left(U\right)=\dfrac{S_{\textrm{ext}}}{2\pi}\int\dfrac{Q \widehat{\Psi}\left(Q\right) \exp\left(iUQ\right)}{iV-Q\left(Q^2+1\right)}dQ\ ,
\label{1D:Profile}
\end{equation}
where we recover the statement above that the dimensionless load $S_{\textrm{ext}}$ gives only a constant prefactor in the linear response.

In order to account for the finite-size effects, we introduce a Lorentzian pressure distribution:
\begin{align}
\Psi\left(U\right) &=\dfrac{\sqrt{Bo}}{\pi\left(U^2+Bo\right)}\ , \notag \\
\widehat{\Psi}\left(Q\right) &=\exp\left(-\sqrt{Bo}\vert Q\vert\right)\ .
\label{1D:Lorentz}
\end{align}
Note that $\int dU\psi=1$, by construction, and that in the limit $Bo \rightarrow 0$ this pressure distribution becomes $\Psi\left(U\right)=\delta\left(U\right)$, or $\widehat{\Psi}\left(Q\right)=1$, where $\delta$ denotes the Dirac distribution.

The associated normalized surface profile is shown in Fig.~\ref{Fig:1DProfiles}, for different values of the reduced speed $V$ and the Bond number $Bo$. At low speed, the deformation profile is found to be nearly symmetric. In that case, the depression is centred at the maximum of the pressure field and it exponentially relaxes towards a flat horizontal surface, far away from the perturbation. As we increase the speed $V$, the symmetry of the surface shape breaks and the depression is shifted towards the rear (left-hand side of Fig.~\ref{Fig:1DProfiles}) of the driving pressure field, while a bump at the front (right-hand side of Fig.~\ref{Fig:1DProfiles}) grows with the disturbance speed. An exponential relaxation is still observed in the rear, with a shorter characteristic length scale as the speed increases, whilst either a similar relaxation (for intermediate speeds) or an exponentially decaying oscillation (for high speeds) is discerned at the front.

\begin{figure}
\centering
\includegraphics[width=0.55\textwidth]{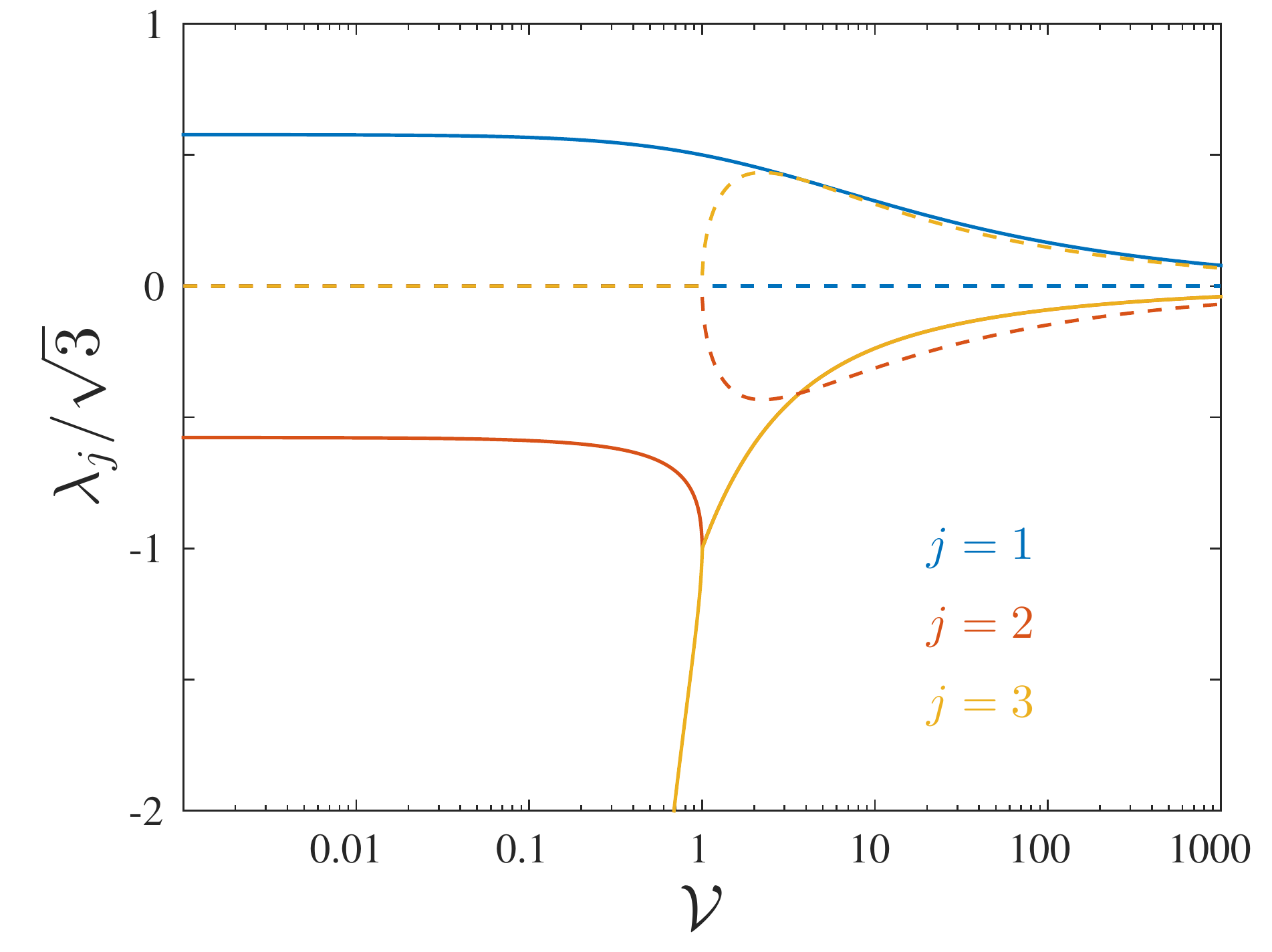}
\caption{Real (plain) and imaginary (dashed) parts of the three dimensionless wavelengths $\lambda_j$ of the asymptotic solution, as given by eq.~(\ref{1D:Sigma}) for the 1D case, as a function of the speed ratio $\mathcal{V}=V/V_{\textrm{c}}=3\sqrt{3}V/2$, where $V$ is the reduced speed. Note that for $\mathcal{V}\leq 1$, $\Im\left(\lambda_j\right)=0$ for all the values of $j$, whereas for $\mathcal{V}\geq 1$, $Re\left(\lambda_2\right)$ and $Re\left(\lambda_3\right)$ overlap.}
\label{Fig:1DLam}
\end{figure} 

These features can be understood by looking at the asymptotic solution of the 1D-TFE. At a sufficiently far position ($\left\vert U\right\vert\gg \sqrt{Bo}$) from the pressure field, its overall effect can thus be taken as an interior boundary condition in order to analyse the spatial spectrum of the solution in the far field. In that approximation, eq.~\eqref{TFE:1D} has the three asymptotic basis solutions:
\begin{equation}
\zeta_j\sim\exp\left(\dfrac{U}{\lambda_j}\right)\ ,
\end{equation}
for $j=1,2,3$, where the wavelengths $\lambda_j$ are given by the characteristic equation:
\begin{equation}
\lambda_j^3+\dfrac{1}{V}\left(\lambda_j^2-1\right)=0\ ,
\end{equation}
and thus:
\begin{align}
\lambda_j &=\sqrt{3}\left(\dfrac{\Omega}{\sigma_j}+\dfrac{\sigma_j}{\Omega}\right)^{-1}\ , \notag \\
\sigma_j &=\exp\left[i\dfrac{2\pi}{3}\left(j-1\right)\right]\ , \notag \\
\Omega &=\sqrt[3]{\mathcal{V}+\sqrt{\mathcal{V}^2-1}}\ .
\label{1D:Sigma}
\end{align}
Here, $\mathcal{V}$ is the speed ratio $\mathcal{V}=V/V_{\textrm{c}}$ defined from the critical speed $V_{\textrm{c}}=2/(3\sqrt{3})$. The dependence of $\lambda_j$ on $\mathcal{V}$ is depicted in Fig.~\ref{Fig:1DLam}.
Since $\Omega$ comes from an expression which may have different values (the roots of $\Omega^6-2\mathcal{V}\Omega^3+1$) we note that the correct value to be employed in eq.~\eqref{1D:Sigma} satisfies the relation $\left[\Re\left(\Omega\right)\right]^2=1-\left[\Im\left(\Omega\right)\right]^2$ for $\mathcal{V}\leq 1$, and $\Im\left(\Omega\right)=0$ for $\mathcal{V}\geq 1$.
  
According to the signs of $\lambda_j$ in Fig.~\ref{Fig:1DLam}, and in order to avoid any divergence of the profile at $U=-\infty$, it follows that for $U<0$ and $\vert U\vert\gg \sqrt{Bo}$ the shape of the surface is necessarily given by:
\begin{equation}
\zeta=N_1\exp\left(\dfrac{U}{\lambda_1}\right)\ ,
\end{equation}
where $N_1$ is a real constant and $\lambda_1$ takes a real value $\Re(\lambda_1)>0$. This indicates that the surface always presents a pure exponential relaxation towards a flat horizontal profile at the rear of the perturbation, in the far field, no matter the speed.
The typical distance $\lambda_1$ over which this relaxation occurs becomes shorter as $\mathcal{V}$ increases, as observed in Fig.~\ref{Fig:1DProfiles}.

On the other hand, for $U>0$ and $\vert U\vert\gg \sqrt{Bo}$, the behaviour of the asymptotic solution depends strongly on $\mathcal{V}$. When $\mathcal{V}<1$, according to the signs of $\lambda_j$ in Fig.~\ref{Fig:1DLam}, and in order to avoid any divergence of the profile at $U=+\infty$, the surface profile is described in the far field by the superposition:
\begin{equation}
\zeta=N_2\exp\left(\dfrac{U}{\lambda_2}\right)+N_3\exp\left(\dfrac{U}{\lambda_3}\right)\ ,
\end{equation}
where $N_2$ and $N_3$ are real constants, and $\lambda_2$ and $\lambda_3$ are both real and negative with $\vert\lambda_2\vert<\vert\lambda_3\vert$.
As a consequence, an over-damping of the surface perturbation is observed in the far field.
The $\mathcal{V}=1$ case corresponds to the critical damping situation for which $\lambda_2=\lambda_3=-1$.
Finally, for $\mathcal{V}>1$, $\lambda_2$ and $\lambda_3$ are complex conjugates, and so are the corresponding amplitudes $N_2$ and $N_3$. Thus, one can write:
\begin{equation}
  \zeta=N\exp\left[\dfrac{\Re\left(\lambda_2\right)U}{\Re\left(\lambda_2\right)^2+\Im\left(\lambda_2\right)^2}\right]\cos\left[\dfrac{\Im\left(\lambda_2\right)U}{\Re\left(\lambda_2\right)^2+\Im\left(\lambda_2\right)^2}+\phi\right]\ ,
\end{equation}
where $N$ and $\phi$ are real constants. This indicates the presence of spatial oscillations at the front of the perturbation in the far field, before the film surface attains its complete relaxation. In brief, the critical speed $V_{\textrm{c}}$ denotes the reduced speed $V$ at which a transition is observed from overdamped relaxation to damped oscillations, at the front of the perturbation. 

In addition to the far-field trends described above, that depend solely on the reduced speed $V$ and the associated dimensionless characteristic lengths $\lambda_j$, one can discern in Fig.~\ref{Fig:1DProfiles} different near-field behaviours of the surface profiles that involve a second dimensionless length $\sqrt{Bo}$ associated with the horizontal extent of the pressure field. As already explained above, for a pressure distribution with sufficiently small size ($\sqrt{Bo}\ll1$), the healing length of the profile is almost independent of $Bo$, and approximately given by $\lambda_j$ that decreases as $V$ increases. In contrast, for a large size ($\sqrt{Bo}\gg1$), the lateral extent of the profile is comparable with the typical size $\sqrt{Bo}$ of the pressure disturbance, as can be observed at low speed $V$ in Fig.~\ref{Fig:1DProfiles}. In this $\sqrt{Bo}\gg1$ case, the healing length decreases with increasing speed only for low velocities. Above the critical speed $V_{\textrm{c}}$ typically, the healing length saturates, leading to surface profiles that nearly follow the same trend, independently of $V$.

\subsection{Wave resistance}
\label{Sec:1DWave}

\begin{figure}
\centering
\includegraphics[width=0.55\textwidth]{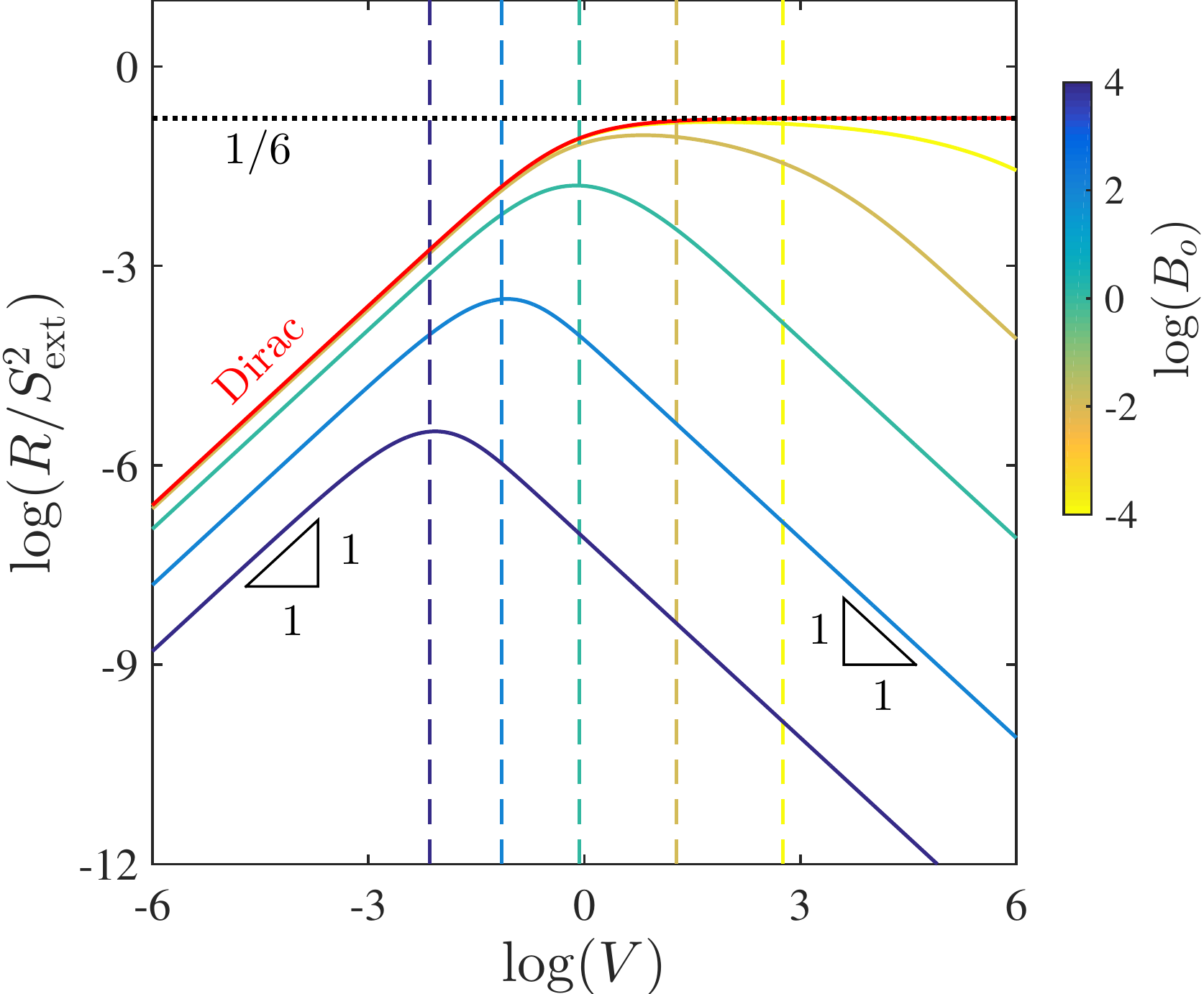}
\caption{Dimensionless wave resistance $R$ (see eq.~\eqref{1D:WRint}) normalized by the square of the dimensionless load $S_{\textrm{ext}}$, as a function of the reduced speed $V$, for a 1D Dirac pressure distribution, and for a 1D Lorentzian pressure distribution (see eq.~\eqref{1D:Lorentz}) with different values of the Bond number $Bo$ as indicated. The vertical dashed lines indicate the crossover speeds $V_{\textrm{s}}$, estimated from eq.~(\ref{1D:WRVs}) for different $Bo$. The horizontal dotted line corresponds to the plateau value of $1/6$ in the 1D Dirac case.}
\label{Fig:1DWR}
\end{figure}

\begin{figure}
\centering
\includegraphics[width=0.45\textwidth]{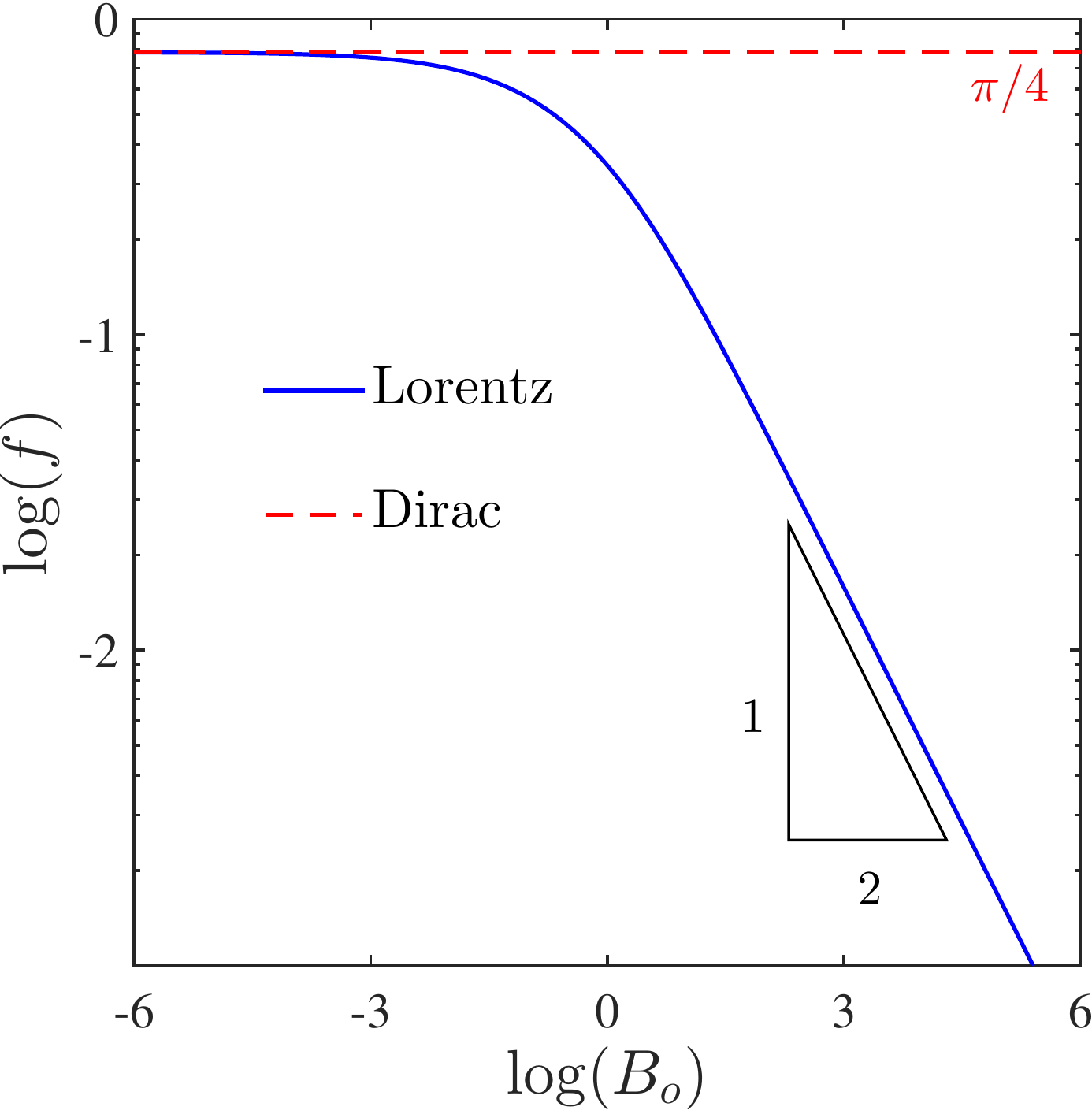}
\caption{Shape function $f$ (see eq.~(\ref{1D:WRzero})) of the 1D pressure field in the low-speed regime, as a function of the Bond number $Bo$. Both the Lorentzian (plain, see eq.~(\ref{1D:fBo})) and the Dirac (dashed, see text) cases are represented.}
\label{Fig:1DfBo}
\end{figure}

Invoking eq.~\eqref{WR:Have} and the fact that in 1D one has $\ell=L_y$, one gets the following 1D expression of the dimensionless wave resistance:
\begin{equation}
R=S_{\textrm{ext}}\int \Psi\left(U\right) \dfrac{\textrm{d}\zeta}{\textrm{d}U} dU\ .
\label{1D:Have}
\end{equation}
For a viscous thin film, it can be shown that the power exerted by the wave resistance equals the viscous dissipated power in the bulk (see App.~\ref{App:1DWR}, eq.~\eqref{powbal}). This power balance highlights the fact that the wave resistance is the force externally furnished by the operator in order to maintain the stationary wake. Remarkably, due to the nature of the TFE, the wave resistance is also related to the spatial energy spectral density (see App.~\ref{App:1DVD}, eq.~\eqref{powspec}).

Combining eqs.~\eqref{1D:Profile} and~\eqref{1D:Have} leads to:
\begin{equation}
R=\dfrac{S_{\textrm{ext}}^2 V}{2\pi}\int\dfrac{Q^2\, \widehat{\Psi}\left(Q\right)\widehat{\Psi}\left(-Q\right)}{V^2+Q^2\left(Q^2+1\right)^2}dQ\ .
\label{1D:WRint}
\end{equation}
Figure \ref{Fig:1DWR} displays the normalized wave resistance for a Lorentzian pressure distribution, as given by eqs.~\eqref{1D:Lorentz} and~\eqref{1D:WRint}, as a function of the reduced speed $V$, and for different values of the Bond number $Bo$. The wave resistance for a Dirac pressure distribution is also shown.

For the asymptotic case where $V\rightarrow 0$, a series expansion of the integrand in eq.~\eqref{1D:WRint} leads to:
\begin{equation}
R=\dfrac{S_{\textrm{ext}}^2 V}{\pi}\left[f\left(Bo\right)+O\left(V^2\right)\right]\ ,
\label{1D:WRzero}
\end{equation}
where $f\left(Bo\right)$ is a function that depends only on the shape of the pressure field. This points out the linear dependence of the wave resistance on $V$, in the small  speed regime, as observed in Fig.~\ref{Fig:1DWR}. For a Dirac pressure distribution one gets $f_{\delta}\left(Bo\right)=\pi/4$, whereas for a Lorentzian (see eq.~(\ref{1D:Lorentz})), one has:
\begin{equation}
\label{1D:fBo}
f_\textrm{L}\left(Bo\right)=8Bo^{3/2}\int_{0}^{\infty}\dfrac{\exp\left(-w\right)}{\left(w^2+4Bo\right)^2}dw\ .
\end{equation}
Both behaviours are depicted in Fig.~\ref{Fig:1DfBo}.
In addition, $f_\textrm{L}\rightarrow\pi/4$ when $Bo\rightarrow0$, as for the Dirac pressure distribution, whilst a series expansion at $Bo\rightarrow\infty$ yields:
\begin{equation}
f_{\textrm{L}}\left(Bo\right)=\dfrac{1}{2\sqrt{Bo}}\left[1+O\left(Bo^{-1}\right)\right]\ ,
\end{equation}
as observed in Fig.~\ref{Fig:1DfBo}.

Considering the Dirac delta distribution, for any $V$, one finds (see App.~\ref{App:1DWR} for details):
\begin{equation}
R_{\delta}=\dfrac{-i \sqrt{3}\theta^{3/2}S_{\textrm{ext}}^2 V}{2\left(\theta^3-2\theta^2+2\theta-1\right)}\ ,
\label{1D:WRDirac}
\end{equation}
with:
\begin{equation}
\theta=\sqrt[3]{1-2\mathcal{V}^2+2\mathcal{V}\sqrt{\mathcal{V}^2-1}}\ ,
\end{equation}
where $\mathcal{V}$ was defined after eq.~\eqref{1D:Sigma}. For the asymptotic case where $V\rightarrow \infty$, a series expansion leads to:
\begin{equation}
R_{\delta}=\dfrac{S_{\textrm{ext}}^2}{6}\left[1+O\left(V^{-2/3}\right)\right]\ .
\end{equation}
This result indicates the singular feature that for a 1D Dirac pressure distribution the wave resistance saturates to a constant value at high speed, as observed in Fig.~\ref{Fig:1DWR}. 

For a finite-size Lorentzian pressure distribution, we combine eqs.~\eqref{1D:Lorentz} and \eqref{1D:WRint} (see App.~\ref{App:1DWR} for details on the overall expression) and obtain in the high-speed regime $V\rightarrow \infty$:
\begin{equation}
R_{\textrm{L}}=\dfrac{S_{\textrm{ext}}^2}{4\pi V Bo^{3/2}}\left[1+O\left(V^{-2}\right)\right]\ .
\end{equation}
Therefore, a $1/V$ dependency of the wave resistance is asymptotically expected for finite-size pressure fields at large speed, as indeed observed in Fig.~\ref{Fig:1DWR}. 

For a Dirac pressure distribution, the reduced speed $V_{\textrm{s}}$ at which the wave resistance saturates can be defined by equating the low- and high-speed asymptotic expressions of $R_{\delta}$. This provides $V_{\textrm{s}}=2/3$, consistent with the observation of Fig.~\ref{Fig:1DWR}. Similarly, for a finite-size Lorentzian pressure distribution, the reduced speed $V_{\textrm{s}}$ at which the crossover between the low- and high-speed regimes of $R_{\textrm{L}}$ occurs is defined by:
\begin{equation}
V_{\textrm{s}}\sim\dfrac{1}{2Bo^{3/4}\sqrt{f(Bo)}}\ ,
\label{1D:WRVs}
\end{equation}
as observed in Fig.~\ref{Fig:1DWR}. In the case of small Bond number $Bo$, $V_{\textrm{s}}$ scales as $Bo^{-3/4}$; whereas for large values of $Bo$, $V_{\textrm{s}}$ scales as $Bo^{-1/2}$ .

\section{Two-dimensional case}
\label{Sec:2D}

\subsection{Surface profile}
\label{Sec:2DSurf}

\begin{figure}
\centering
\includegraphics[width=0.55\textwidth]{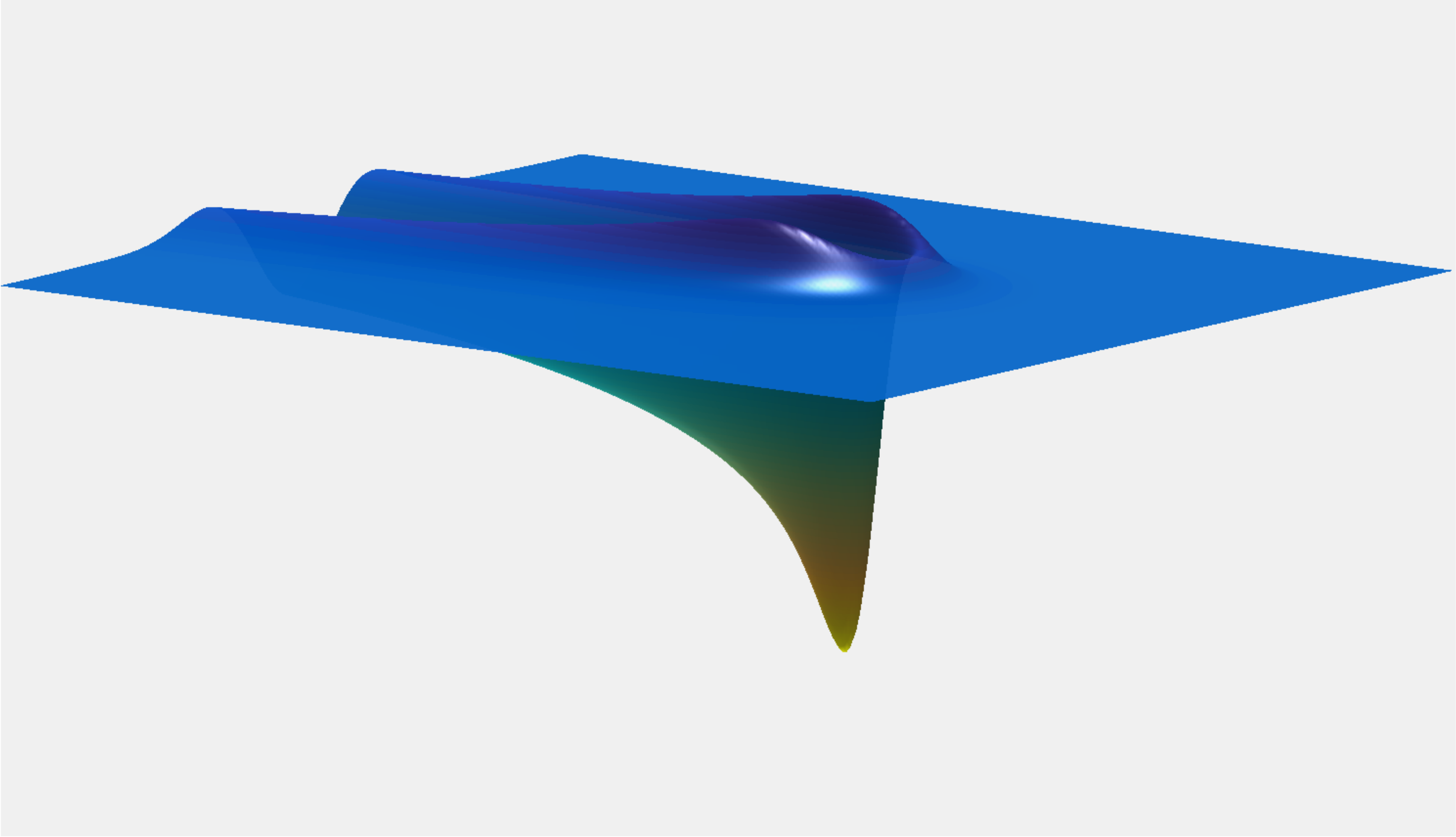}
\includegraphics[width=0.55\textwidth]{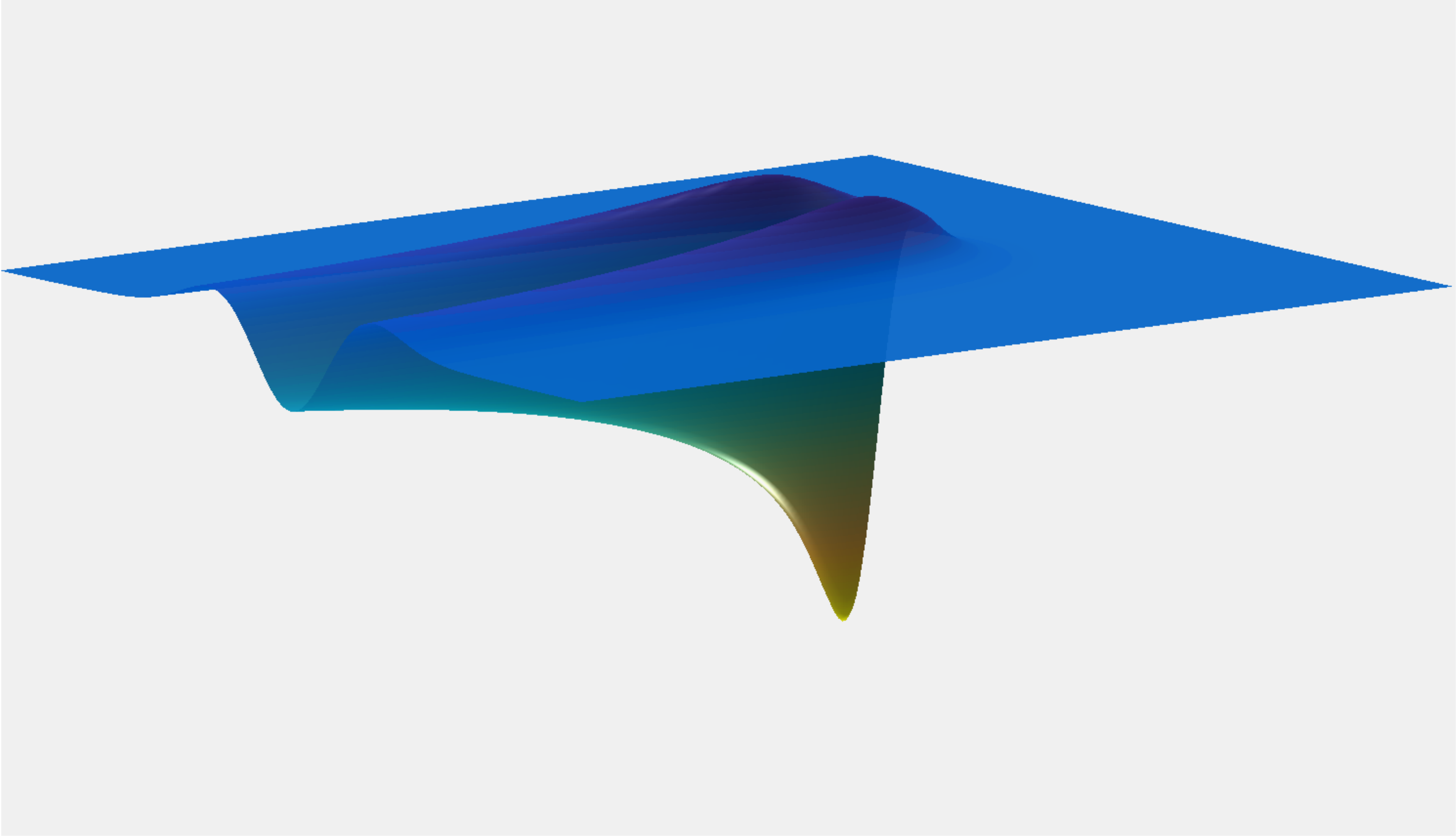}
\caption{Normalized thin-film surface profile in presence of a moving Lorentzian pressure disturbance, with dimensionless load $S_{\textrm{ext}}>0$, Bond number $Bo=10^2$, and reduced speed $V=10$, computed from eqs.~\eqref{2D:Profile} and \eqref{2D:Lorentz}. Perspective views of the front (top) and the back (bottom), in arbitrary units and colour code.}
\label{Fig:2DSurf}
\end{figure}

\begin{figure}
\centering
\includegraphics[width=0.87\textwidth]{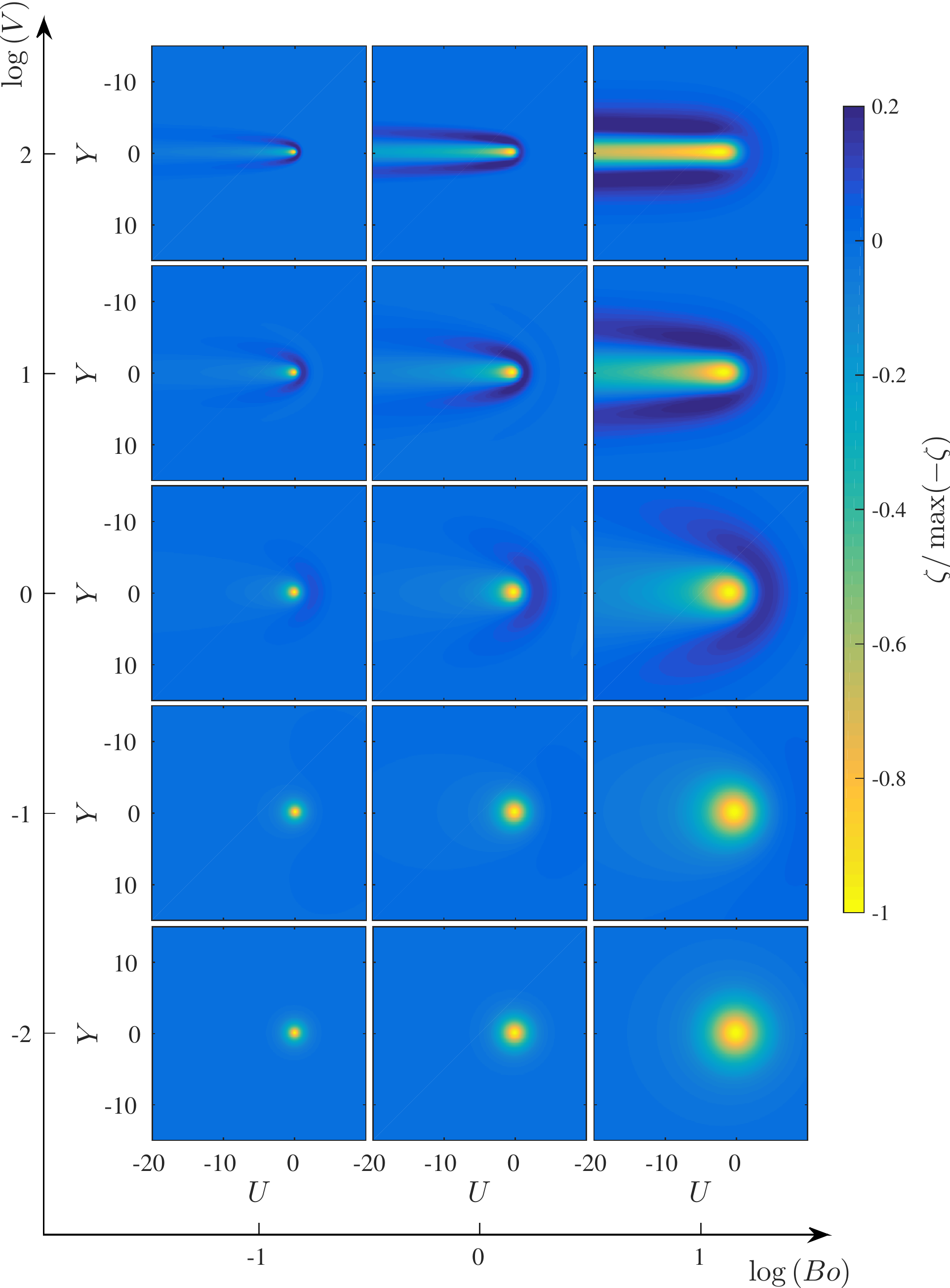}
\caption{Normalized thin-film surface profile (top view) computed from eqs.~\eqref{2D:Profile} and \eqref{2D:Lorentz}, for a dimensionless load $S_{\textrm{ext}}>0$, and different values of the Bond number $Bo$ and the reduced speed $V$. In each subfigure, the pressure field travels from left to right, along the longitudinal coordinate $U$.}
\label{Fig:2DProfiles}
\end{figure}

Considering the two-dimensional Fourier transforms $\widehat{\zeta}\left(Q,K\right)$ and $\widehat{\Psi}\left(Q,K\right)$ of the surface profile and the pressure field, with the definitions given in App.~\ref{App:2DFT}, the 2D-TFE of eq.~\eqref{TFE:2D} becomes:
\begin{equation}
\left[iQV-\left(Q^2+K^2\right)^2-\left(Q^2+K^2\right)\right]\widehat{\zeta}=S_{\textrm{ext}}\left(Q^2+K^2\right)\widehat{\Psi}\ ,
\label{2D:TFEFT}
\end{equation}
where $Q$ and $K$ are the spatial angular frequencies in the $U$ and $Y$-directions, respectively.
Consequently, the surface $\zeta\left(U,Y\right)$ of the thin film reads:
\begin{equation}
\zeta\left(U,Y\right)=\dfrac{S_{\textrm{ext}}}{4\pi^2}\iint\dfrac{\left(Q^2+K^2\right) \exp\left[i\left(QU+KY\right)\right]\widehat{\Psi}\left(Q,K\right)}{iQV-\left(Q^2+K^2\right)\left(1+Q^2+K^2\right)}dKdQ\ .
\label{2D:Profile}
\end{equation}

In analogy with the previous 1D case, we introduce the following 2D axisymmetric finite-size Lorentzian pressure distribution:
\begin{align}
\Psi\left(U,Y\right) &=\dfrac{\sqrt{Bo}}{2\pi\left(U^2+Y^2+Bo\right)^{3/2}}\ , \notag \\
\widehat{\Psi}\left(Q,K\right)&=\exp\left[-\sqrt{Bo\left(Q^2+K^2\right)}\right]\ .
\label{2D:Lorentz}
\end{align}
We recall that, since $a$ is the characteristic horizontal size of the pressure field, $\sqrt{Bo}=a \kappa$ is the dimensionless size. When, $Bo\rightarrow 0$, the Dirac pressure distribution $\Psi\left(U,Y\right)=\delta(U)\delta(Y)$, or its equivalent $\widehat{\Psi}\left(Q,K\right)=1$ in the frequency domain,  is recovered. 

Surface patterns generated by a Lorentzian pressure disturbance, for several combinations of the parameters $Bo$ and $V$, are illustrated in Figs.~\ref{Fig:2DSurf} and~\ref{Fig:2DProfiles}. A common feature of these profiles is the depression observed. In Fig.~\ref{Fig:2DSurf}, due to the particular parameters employed (see caption) to generate the surface profile, a non-symmetric shape is shown. A rim surrounds the front of the depression and becomes a wake, behind, that follows the straight translation path of the pressure distribution along $U$. 

In Fig.~\ref{Fig:2DProfiles}, for low speeds ($V<1$), we find deformation profiles which are mostly symmetric, with a main depression aligned with the maximum of the pressure field and followed by a monotonic relaxation towards a flat horizontal surface. In this situation, the liquid-film surface relaxation occurs faster than the displacement of the pressure field: in real units, the displacement speed $v$ is smaller than the characteristic speed $\rho g h_0^3 \kappa/3\mu$ at which the film relaxes and thus propagates information at its surface. As the pressure acts on the surface by creating a moving depression, the film has enough time to move the dislodged liquid volume isotropically. 

Like for the 1D case, an increase of the speed $V$ induces a symmetry breaking phenomenon, which above $V\simeq 1$ consists of a level rising at the front (right-hand side of the corresponding subfigures of Fig.~\ref{Fig:2DProfiles}) and a wake at the rear (left-hand side). In real units, the relaxation of the liquid film now occurs with a speed $\rho g h_0^3 \kappa/3\mu$ which is smaller than the displacement speed $v$ of the pressure field. In other words, the liquid volume that is ejected by the applied pressure is not distributed isotropically because the information at the film surface does not propagate fast enough. This retardation effect is similar to the Mach and Cerenkov physics~\cite{Carusotto2013}, where the speed of the object becomes larger than the wave-propagation speed in the considered medium.

For relatively moderate speeds, $V\simeq 1$, the protuberance at the front covers a crescent-shaped region that grows with the size of the pressure field $\sqrt{Bo}$.
The maximum surface level at the front of the pressure field is located at $Y=0$, whereas the maximum of the wake, at each position $U<0$, is located at $Y\sim\left\vert U\right\vert^{1/2}$.
In contrast, the local minimum of the wake is always placed at $Y=0$.
In addition, both the local maximum and minimum of the wake present a rapid spatial exponential relaxation.
As the speed grows in the range $V\in \left[1,10^2\right]$, the height of the frontal rim reduces, while the extent of the wake in the rear reaches a larger distance.
The local maximum at the front is still located at $Y=0$, while the maximum height of the wake is now placed at $Y\sim \left\vert U\right\vert^{\alpha}$, where the exponent $\alpha$ evolves from $1/2$ at $V\sim 1$ to $1/4$ at $V\sim 10^2$.
In fact, the $1/4$ exponent in the high-speed regime indicates the same dynamics as the capillary leveling of a trench in a viscous thin film with no slip at the substrate~\cite{Baumchen2013,BenzaquenSM2014}.
In this regime, the film seems to overlook the presence of the fast disturbance, perceiving the footprint left by the latter as a 1D trench which the film tends to refill by capillary leveling.
Therefore, when we compare the transverse profiles at different positions $U<0$, we in fact discern different stages of the 1D leveling of a viscous film.

\subsection{Wave resistance}
\label{Sec:2DWave}

\begin{figure}
\centering
\includegraphics[width=0.55\textwidth]{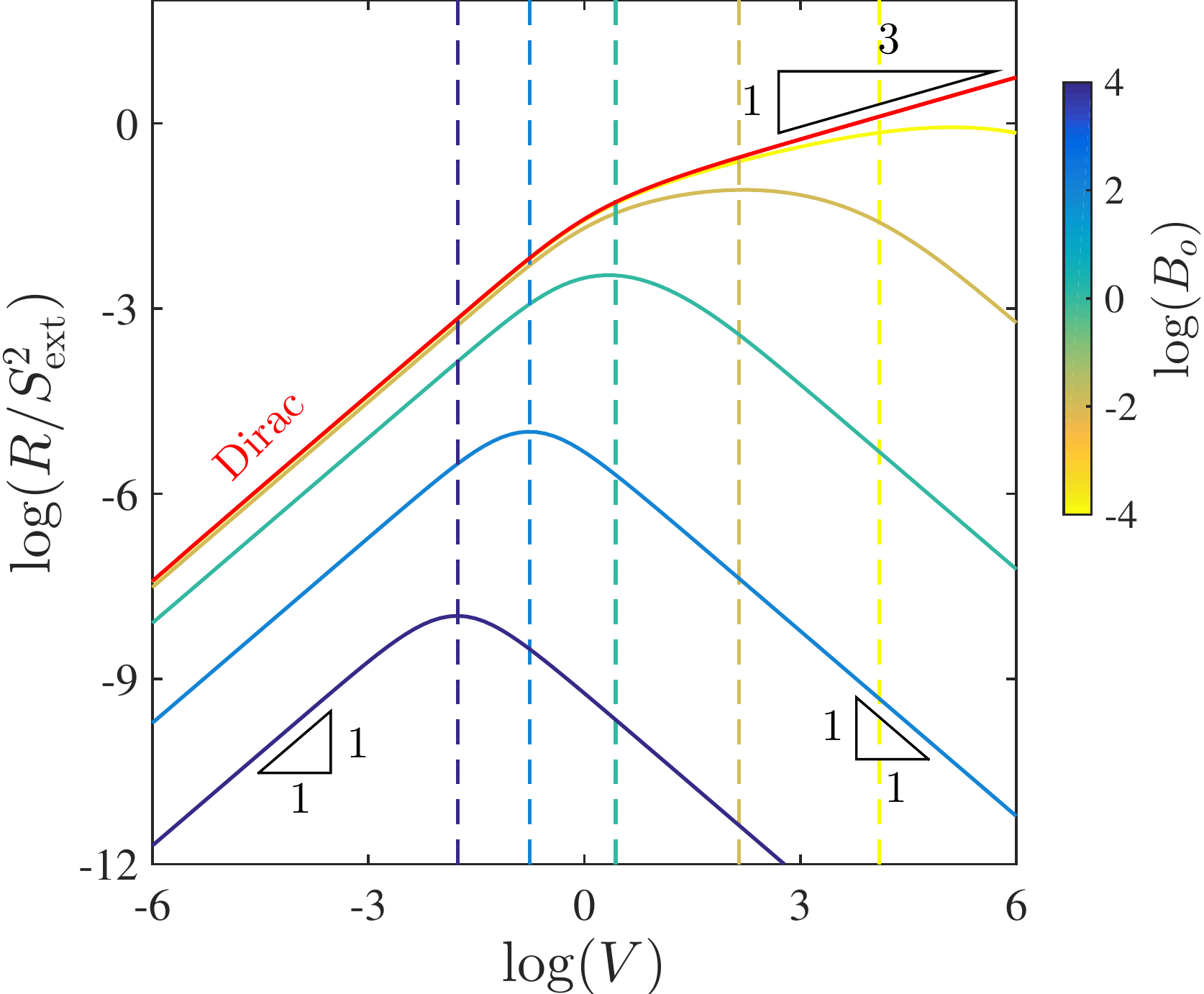}
\caption{Dimensionless wave resistance $R$ normalized by the square of the dimensionless load $S_{\textrm{ext}}$, as a function of reduced speed $V$, for a 2D Dirac pressure distribution (see eq.~\eqref{2D:WRDirac}), and for a 2D Lorentzian pressure distribution (see eqs.~\eqref{2D:Lorentz} and \eqref{2D:WRaxi}) with different values of the Bond number $Bo$ as indicated. The vertical dashed lines indicate the crossover speeds $V_{\textrm{s}}$, estimated from eq.~(\ref{2D:WRVs}) for different $Bo$.}
\label{Fig:2DWR}
\end{figure}

\begin{figure}
\centering
\includegraphics[width=0.45\textwidth]{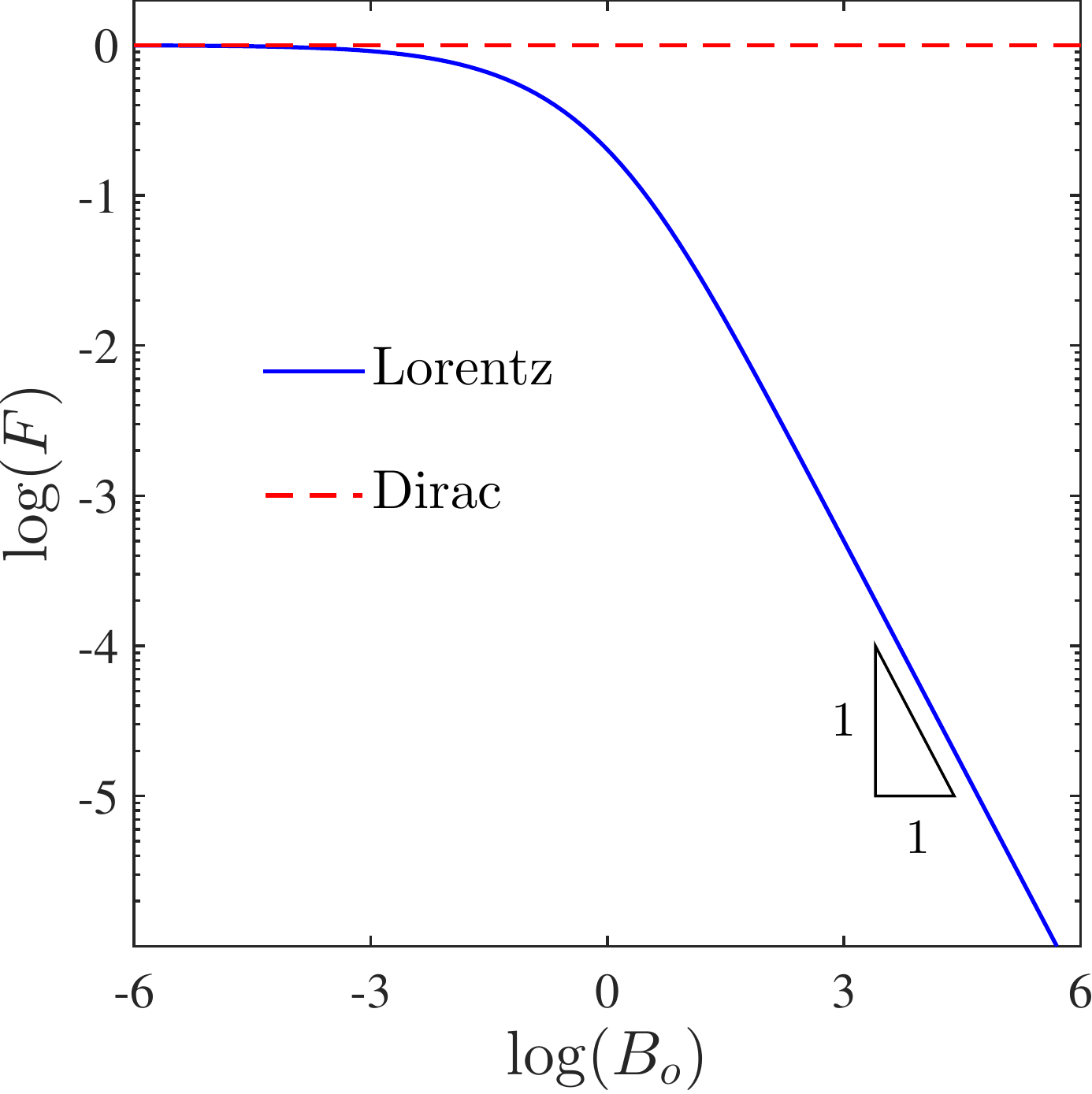}
\caption{Shape function $F$ (see eq.~(\ref{2D:WRzero})) of the 2D pressure field in the low-speed regime, as a function of the Bond number $Bo$. Both the Lorentzian (plain, see eq.~(\ref{2D:FBo})) and the Dirac (dashed, see text) cases are represented.}
\label{Fig:2DFBo}
\end{figure}

Invoking eq.~\eqref{WR:Have} and the fact that in 2D one has $\ell=\kappa^{-1}$, one gets the following 2D expression of the dimensionless wave resistance:
\begin{equation}
R=S_{\textrm{ext}}\iint \Psi\left(U,Y\right) \dfrac{\partial \zeta}{\partial U} dU dY\ .
\label{2D:Have}
\end{equation}
Therefore, in combination with eq.~\eqref{2D:Profile}, one obtains:
\begin{equation}
R=\dfrac{S_{\textrm{ext}}^2V}{4\pi^2}\iint\dfrac{Q^2\left(Q^2+K^2\right)\widehat{\Psi}\left(Q,K\right)\widehat{\Psi}\left(-Q,-K\right)}{Q^2V^2+\left(Q^2+K^2\right)^2\left(1+Q^2+K^2\right)^2}\, dK\,dQ\ .
\label{2D:WRint}
\end{equation}

Considering the polar coordinates $\left(\varrho,\varphi\right)$, an axisymmetric pressure field $\widehat{\Psi}\left(\varrho\right)$, as it is the case for the Dirac and Lorentzian distributions presented in this study, and performing an integration over $\varphi$ (see App.~\ref{App:2DWR}) yields:
\begin{equation}
R=\dfrac{S_{\textrm{ext}}^2}{2\pi V}\int^{\infty}_{0}\left[1-\dfrac{\varrho\left(\varrho^2+1\right)}{\sqrt{V^2+\varrho^2\left(\varrho^2+1\right)^2}}\right]\left[\widehat{\Psi}\left(\varrho\right)\right]^2 \varrho^3 d\varrho\ .
\label{2D:WRaxi}
\end{equation}
For a Lorentzian distribution, a combination of eqs.~\eqref{2D:Lorentz} and \eqref{2D:WRaxi} leads to the corresponding wave resistance, which behaviour in terms of the reduced speed $V$ is shown in Fig.~\ref{Fig:2DWR}, for different values of $Bo$.
The wave resistance for a Dirac distribution is also represented.

In the asymptotic case where $V\rightarrow 0$, a series expansion of the integrand in eq.~\eqref{2D:WRaxi} (see App.\ref{App:2DWR} for details) leads to:
\begin{equation}
R=\dfrac{S_{\textrm{ext}}^2 V}{8\pi}\left[F\left(Bo\right)+O\left(V^2\right)\right]\ ,
\label{2D:WRzero}
\end{equation}
where $F\left(Bo\right)$ is a function that depends only on the shape of the pressure field, pointing out the linear dependency of the wave resistance on $V$, in the low reduced-speed regime.
For a Dirac pressure field $F_{\delta}\left(Bo\right)=1$, whereas for a Lorentzian, it is described by:
\begin{equation}
\label{2D:FBo}
F_\textrm{L}\left(Bo\right)=8Bo\int_{0}^{\infty}\dfrac{\exp\left(-w\right)w}{(w^2+4 Bo)^2}dw\ .
\end{equation}
Both trends are illustrated in Fig.~\ref{Fig:2DFBo}.
For the Lorentzian $F_\textrm{L}\left(Bo\right)=1$ at $Bo\ll 1$,  the same behavior as for the Dirac pressure distribution, while a series expansion at $Bo\rightarrow\infty$ gives:
\begin{equation}
F_{\textrm{L}}\left(Bo\right)= \dfrac{1}{2Bo}\left[1+O\left(Bo^{-1}\right)\right]\ .
\end{equation}

Coming back to the Dirac pressure distribution, a series expansion of eq.~\eqref{2D:WRaxi} at $V\rightarrow\infty$ yields:
\begin{equation}
R_{\delta}=\dfrac{\Gamma\left(\frac{1}{3}\right)\Gamma\left(\frac{7}{6}\right)}{8\pi^{3/2}}S_{\textrm{ext}}^2 V^{1/3}\left[1+O\left(V^{-1}\right)\right]\ ,
\label{2D:WRDiracInf}
\end{equation}
which indicates that the wave resistance of a 2D Dirac pressure diverges slowly with increasing $V$.

For a Lorentzian pressure distribution, a series expansion in the high-speed regime where $V\rightarrow\infty$ allows us to reduce the combination of eqs.~\eqref{2D:WRaxi} and \eqref{2D:Lorentz} into:
\begin{equation}
R_{\textrm{L}}=\dfrac{3S_{\textrm{ext}}^2}{16\pi V Bo^2}\left[1+O\left(V^{-2}\right)\right]\ .
\label{2D:WRLorenzInf}
\end{equation}
This demonstrates that, for finite-size pressure fields in the high-speed regime where $V\rightarrow\infty$, the wave resistance behaves in the same way as in the 1D case, \textit{i.e.} decreasing as $1/V$.

For a Dirac distribution, the speed dependence of the wave resistance changes from $R\sim V$ to $R\sim V^{1/3}$ at the reduced speed $V_{\textrm{s}}=5/3$, which is obtained by equating the low- and high-speed asymptotic expressions of $R_{\delta}$.
Similarly, for a finite-size pressure distribution, the reduced speed $V_{\textrm{s}}$, at which the crossover between the two asymptotic regimes takes place, reads:
\begin{equation}
V_{\textrm{s}}\sim\sqrt{\dfrac{3}{2Bo^2 F\left(Bo\right)}}\ ,
\label{2D:WRVs}
\end{equation}
emerging when we match the low- and high-speed regimes of $R_{\textrm{L}}$.
At small Bond number $Bo$, the crossover speed is approximately given by $V_{\textrm{s}}\sim\left(Bo\sqrt{2/3}\right)^{-1}$, whilst for large values of $Bo$, it happens around $V_{\textrm{s}}\sim\left(Bo/3\right)^{-1/2}$.

\section{Conclusion}

In the present work, the effects of a pressure disturbance moving near the free surface of a viscous thin film have been studied in detail.
Depending on the disturbance-speed regime, the existence of nearly symmetric and non-symmetric profile shapes have been revealed by means of numerical methods.
The nature of the wakes and bumps observed at the surface, as well as the lifespan of the trail and its relaxation towards a flat surface in the far field, have been explained through asymptotic solutions, unveiling their reliance on several dimensionless parameters: the Bond number $Bo$, the dimensionless load $S_{\textrm{ext}}$ and the dimensionless speed of the disturbance $V$, which are related to the geometrical and physical properties of the liquid film: the density $\rho$, the air-liquid surface tension $\gamma$, the dynamic viscosity $\mu$, the film thickness $h_0$ and the pressure distribution through its total load $s_{\textrm{ext}}$ and width $a$.

In addition, the wave resistance, which one must overcome in order to displace the disturbance at constant speed, has been analyzed.
Even though the shape of the free surface is calculated numerically, the wave resistance can be derived in different asymptotic speed regimes, by means of analytical expressions for Dirac delta and Lorentzian pressure distributions.
Moreover, a general expression of the wave resistance for a Dirac delta distribution in the 1D case has been presented.

We believe that the measurement of the wave resistance can be employed to deduce one of the physical properties of a viscous thin film, such as the above $\rho,\ \gamma,\ \mu$ or $h_0$, when the remaining parameters are known.
The working principle of a new fine rheological probe may be based on this acquired insight.
Additionally, in combination with local quenching, controlling the lifespan of the footprint generated by the moving perturbation might serve as the technological foundation of a new nano-patterning technique, which may allow for imprinting nano-channels over polymeric materials.
This idea may provide an alternative to the current nano-lithography techniques employed to fabricate nano-devices and templates for electronics.

\section{Acknowledgements}

We wish to thank D. Legendre, P. Tordjeman and  A. Darmon for fruitful and very interesting discussions.

\footnotesize{
\bibliography{WakeWaveTFE} 
\bibliographystyle{unsrt}
}

\appendix
\section{Details on the one-dimensional case}

\subsection*{Fourier transform}
\label{App:1DFT}

\begin{align}
\widehat{f}\left(Q\right) &=\int_{-\infty}^{\infty}f\left(U\right)\exp\left(-iUQ\right)dU\notag\ ,\\
f\left(U\right) &=\dfrac{1}{2\pi}\int_{-\infty}^{\infty}\widehat{f}\left(Q\right)\exp\left(iUQ\right)dQ\ .
\end{align}

\subsection*{Wave resistance}
\label{App:1DWR}

\begin{figure}
\centering
\includegraphics[width=0.54\textwidth]{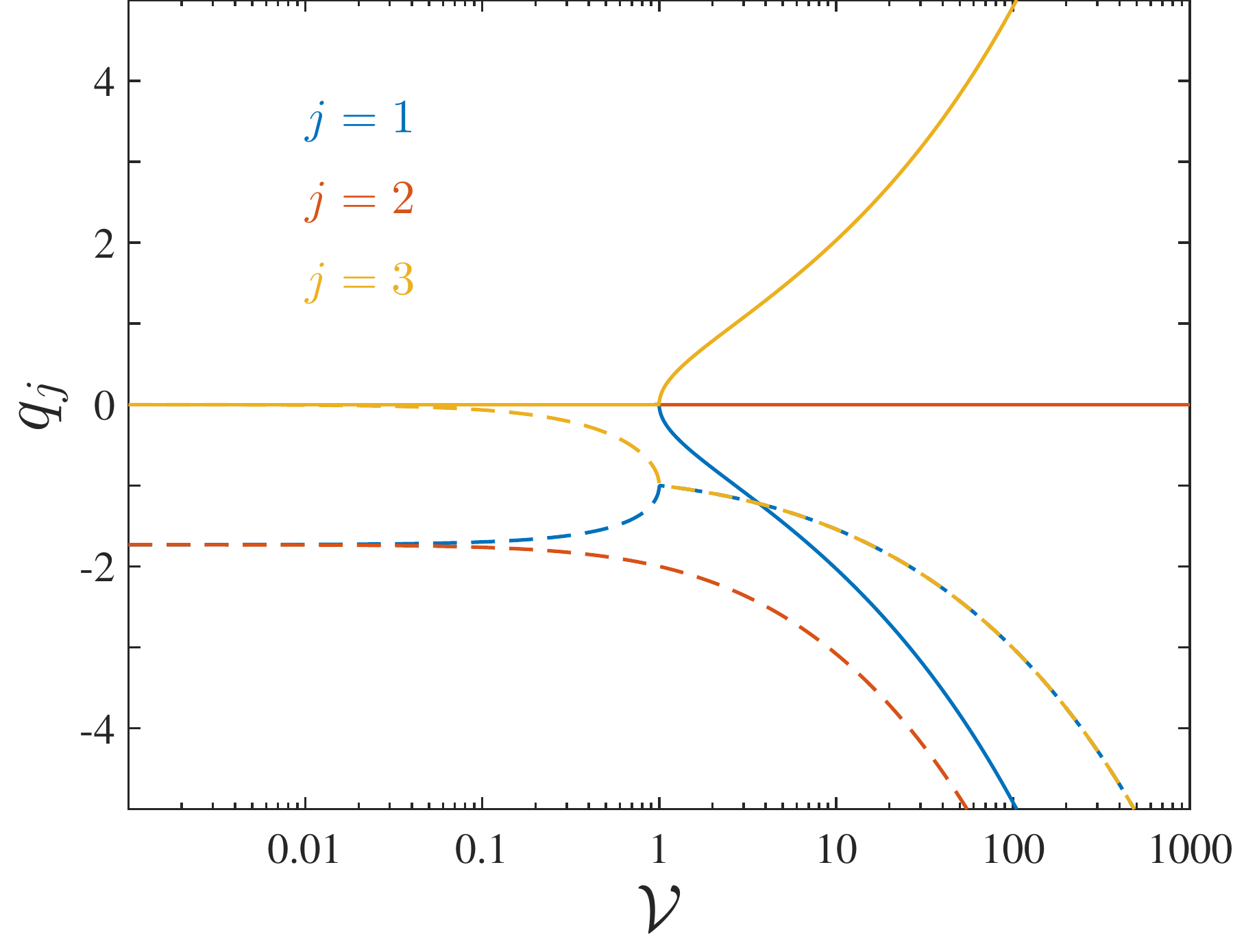}
\caption{Real (plain) and imaginary (dashed) parts of the three dimensionless auxiliary quantities $q_j$ as a function of the speed ratio $\mathcal{V}$, as given by eq.~(\ref{auxi}). Note that for $\mathcal{V}\leq 1$, $\Re\left(q_j\right)=0$ for all the values of $j$, whereas for $\mathcal{V}\geq 1$, $\Im\left(q_1\right)$ and $\Im\left(q_3\right)$ overlap.}
\label{Fig:1DTheta}
\end{figure}

Multiplying eq.~\eqref{TFE:1D} by the surface displacement, and integrating over the whole $U$ domain gives:
\begin{equation}
\int\left(\dfrac{d^3\zeta}{d U^3}-\dfrac{d \zeta}{d U}-V\zeta \right)\zeta d U=S_{\textrm{ext}}\int\left(\dfrac{d\Psi}{d U}\right)\zeta d U\ .
\end{equation}

Some of the terms (the first two from the left-hand side and the one in the right-hand side) of this equation can be simplified to:
\begin{align}
\int\zeta\dfrac{d^3\zeta}{d U^3} d U
  &=\left[\zeta\dfrac{d^2\zeta}{d U^2}-\dfrac{1}{2}\left(\dfrac{d\zeta}{d U}\right)^2\right]_{-\infty}^{\infty}=0\ ,  \notag \\
\int\zeta\dfrac{d\zeta}{d U} d U
  &=\left[\dfrac{1}{2}\left(\zeta\right)^2\right]_{-\infty}^{\infty}=0\ , \notag \\
\int \zeta\dfrac{d\Psi}{d U} d U
  &=\left[\vphantom{\frac{1}{2}}\zeta\Psi\right]_{-\infty}^{\infty}-\int \Psi\dfrac{d \zeta}{d U} d U
  =-\int \Psi\dfrac{d \zeta}{d U} d U\ ,
\end{align}
if one takes into account that for $\vert U \vert\rightarrow\infty$, the surface has completely recovered its flat shape, and thus that $\zeta$ and all its derivatives are equal to zero in the far field. Therefore, one finds:
\begin{equation}
\label{powspec}
S_{\textrm{ext}}\int \Psi\dfrac{d \zeta}{d U} d U=V\int\zeta^2 d U\ ,
\end{equation}
where the left-hand side term is the definition of the wave resistance given in eq.~\eqref{1D:Have} and the integral in the right-hand side is the spatial energy spectral density.

Injecting the Dirac delta pressure distribution into eq.~\eqref{1D:WRint} and performing the integration, yields the following expression of the wave resistance:
\begin{equation}
R_{\delta}=\dfrac{i 3\sqrt{3} S_{\textrm{ext}}^2 V}{2\left(q_1+q_2\right)\left(q_1+q_3\right)\left(q_2+q_3\right)}\ ,
\end{equation}
where $i=\sqrt{-1}$ is the imaginary unit, and where we introduced the auxiliary quantities $q_1$, $q_2$ and $q_3$, which are given by (see Fig.~\ref{Fig:1DTheta}):
\begin{align}
\label{auxi}
q_j &=\left(\dfrac{\theta}{\sigma_j}\right)^{1/2}-\left(\dfrac{\sigma_j}{\theta}\right)^{1/2}\ , \notag \\
\theta &=\left(1-2\mathcal{V}^2+2\mathcal{V}\sqrt{\mathcal{V}^2-1}\right)^{1/3}\ ,
\end{align}
and where the $\sigma_j$ and $\mathcal{V}$ have been introduced in eq.~\eqref{1D:Sigma}.
After some algebra, one is able to regain eq.~\eqref{1D:WRDirac}.

From the substitution of a Lorentzian distribution into eq.~\eqref{1D:WRint} and after integration, the following expression of the wave resistance is obtained:
\begin{equation}
R_{\textrm{L}}=\dfrac{S_{\textrm{ext}}^2 V}{2\pi}\sum_{\substack{j=-3 \\ j\neq 0}}^{3}\dfrac{Q_j}{1+4Q_j^2+3Q_j^4}\int_{0}^{\infty}\dfrac{-\exp\left(-w\right)}{w-2Q_j \sqrt{Bo}}dw\ ,
\label{1D:WRLorentz}
\end{equation}
where $Q_j$ is the $j$th root of $V^2+Q^2\left(Q^2+1\right)^2$, given by:
\begin{equation}
Q_j=\dfrac{1}{\sqrt{3}}\left(\dfrac{\theta}{\sigma_j}+\dfrac{\sigma_j}{\theta}-2\right)^{1/2}\ ,
\end{equation}
for $j=1,2,3$, whereas for $j=4,5,6$ we have $Q_j=-Q_{j-3}$.
$\sigma_j$ and $\theta$ have been introduced in eqs.~\eqref{1D:Sigma} and \eqref{auxi}, respectively.
The asymptotic behaviours for $V\rightarrow \infty$ can also be obtained from eq.~\eqref{1D:WRLorentz}, after making $\theta\rightarrow 1$ and some algebra.

\subsection*{Viscous dissipation}
\label{App:1DVD}

The viscously-dissipated power indicates the rate at which the external work is irreversibly converted into internal energy by viscous forces.
For a thin film, the viscously-dissipated power per unit length is expressed as:
\begin{equation}
  \Phi_{\mu}=\mu \int\left[\int_{-h_0}^{h}\left(\dfrac{\partial \upsilon_x}{\partial z}\right)^2 dz\right]dx\ ,
\end{equation}
where $\upsilon_x$ is the horizontal component of the velocity within the film.
Under the assumptions of the lubrication theory, the previous expression can be re-written as:
\begin{equation}
  \Phi_{\mu}=\dfrac{1}{\mu} \int\left\{\int_{-h_0}^{h}\left[\left(z-h\right)\dfrac{\partial \Delta p}{\partial x}\right]^2 dz\right\}dx\ .
\end{equation}
Integration over the local film thickness and, afterwards, integration by parts yields:
\begin{align}
\Phi_{\mu} &=\dfrac{1}{3\mu} \int\left(h_0+h\right)^3\left(\dfrac{\partial \Delta p}{\partial x}\right)^2 dx\ , \notag \\
&=\int\Delta p \dfrac{\partial}{\partial x}\left[\dfrac{\left(h_0+h\right)^3}{3\mu}\left(\dfrac{\partial \Delta p}{\partial x}\right)\right] dx\ ,
\end{align}
since $h$, $\Delta p$ and their derivatives, with respect to $x$, are equal to zero at $\vert x\vert\rightarrow\infty$.
Making use of the two-dimensional analogue of eq.~\eqref{TFE:Lub}, allows us to find:
\begin{equation}
  \Phi_{\mu}=\int\Delta p \left(\dfrac{\partial h}{\partial t}\right) dx\ .
\end{equation}
Now, invoking the dimensionless variables and placing ourselves in the moving reference frame, one has:
\begin{equation}
  \Phi_{\mu}=-v \rho g h_0^2 \int\Delta P \left(\dfrac{\partial \zeta}{\partial U}\right) dU\ .
\end{equation}
Then, using the dimensionless equivalent of eq.~\eqref{TFE:Young}, one finds:
\begin{equation}
  \Phi_{\mu}=-v \rho g h_0^2 \int\left(-\dfrac{\partial^2\zeta}{\partial U^2}+\zeta-S_{\textrm{ext}}\Psi\right) \dfrac{\partial \zeta}{\partial U} dU\ .
\end{equation}
Once more, considering that $\zeta$ and its first derivative with respect to $U$ are equal to zero at $\vert U \vert\rightarrow\infty$, the first two terms from the right-hand side of this equation are reduced to:
\begin{align}
  \int\dfrac{d\zeta}{d U}\dfrac{d^2\zeta}{d U^2} d U
    &=\left[\dfrac{1}{2}\left(\dfrac{d\zeta}{d U}\right)^2\right]_{-\infty}^{\infty}=0\ ,  \notag \\
  \int\zeta\dfrac{d\zeta}{d U} d U
    &=\left[\dfrac{1}{2}\left(\zeta\right)^2\right]_{-\infty}^{\infty}=0\ .
\end{align}
Finally, using the definition given in eq.~\eqref{1D:Have}, the viscously-dissipated power can be written in terms of the wave resistance:
\begin{equation}
\label{powbal}
  \Phi_{\mu}= \dfrac{\rho g h_0^2}{\tau \kappa} V R=v r\ .
\end{equation}
As already mentioned, the product $v r$ is the power required to displace the surface profile at constant speed, and it also corresponds to the power dissipated within the viscous thin film during such a task.

\section{Details on the two-dimensional case}

\subsection*{Fourier transform}
\label{App:2DFT}

\begin{align}
\widehat{f}\left(Q,K\right) &=\int^{\infty}_{-\infty}\int_{-\infty}^{\infty}f\left(U,Y\right)\exp\left(-i\left[QU+KY\right]\right)dYdU\ , \notag \\
f\left(U,Y\right) &=\dfrac{1}{4\pi^2}\int^{\infty}_{-\infty}\int_{-\infty}^{\infty}\widehat{f}\left(Q,K\right)\exp\left(i\left[QU+KY\right]\right)dKdQ\ ,
\end{align}

\subsection*{Wave resistance}
\label{App:2DWR}

Starting from eq.~\eqref{2D:WRint}, one considers the polar coordinates $\left(\varrho,\varphi\right)$, given by the relations $Q=\varrho\cos\left(\varphi\right)$ and $K=\varrho\sin\left(\varphi\right)$, and multiplies both the numerator and denominator by $i/V$, in order to retrieve the following expression:
\begin{equation}
R=\dfrac{S_{\textrm{ext}}^2}{4\pi^2 V}\int^{\infty}_{0}\int^{2\pi}_{0}\dfrac{\varrho^3\cos\left(\varphi\right)\widehat{\Psi}\left(\varrho,\varphi\right)\widehat{\Psi}\left(\varrho,\varphi+\pi\right)}{\cos\left(\varphi\right)+i\dfrac{\varrho}{V}\left(\varrho^2+1\right)} d\varphi d\varrho\ .
\label{2D:WRpolar}
\end{equation}
Additionally, if the Fourier transform of the pressure field $\widehat{\Psi}$ under consideration is axisymmetric, a first integration of the previous equation over $\varphi$ leads directly to eq.~\eqref{2D:WRaxi}.
In brief, for an axisymmetric pressure field, one should use the following formula:
\begin{align*}
\int_{0}^{2\pi} \dfrac{\cos(\varphi) d\varphi}{\cos(\varphi)+D} &=2\int_{0}^{\pi} \dfrac{\cos(\varphi) d\varphi}{\cos(\varphi)+D}\ , \\
&=2\left\{\dfrac{2D}{\sqrt{1-D^2}}\tanh^{-1}\left[\dfrac{\left(D-1\right)\tan\left(\dfrac{\varphi}{2}\right)}{\sqrt{1-D^2}}\right]+\varphi\right\}\Bigg\vert_{0}^{\pi}\ ,
\end{align*}
which, considering that:
\begin{equation*}
\lim_{x\rightarrow\infty}\left[\tanh^{-1}\left(x\right)\right]=\dfrac{i\pi}{2}\ ,
\end{equation*}
can be re-written as:
\begin{equation*}
\int_{0}^{2\pi} \dfrac{\cos(\varphi) d\varphi}{\cos(\varphi)+D}=2\pi\left(1+\dfrac{i D}{\sqrt{1-D^2}}\right)\ ,
\end{equation*}
in order to perform the aforementioned integration step.
Moreover, a series expansion of the integrand in eq.~\eqref{2D:WRaxi} near $V=0$ leads to:
\begin{align}
R_0 &=\dfrac{S_{\textrm{ext}}^2 V}{2\pi} \left\{\int^{\infty}_{0} \left[\widehat{\Psi}\left(\varrho\right)\right]^2 \dfrac{\varrho d\varrho}{2\left(\varrho^2+1\right)^2}+O\left(V^2\right)\right\}\ ,
\end{align}
which can be re-written in the form of eq.~\eqref{2D:WRzero}.

For a Dirac pressure distribution, eq.~\eqref{2D:WRaxi} becomes:
\begin{equation}
R_{\delta}=\dfrac{S_{\textrm{ext}}^2}{2\pi V}\int^{\infty}_{0}\left[1-\dfrac{\varrho\left(\varrho^2+1\right)}{\sqrt{V^2+\varrho^2\left(\varrho^2+1\right)^2}}\right] \varrho^3 d\varrho\ .
\label{2D:WRDirac}
\end{equation}
A series expansion near $V=0$ of the integrand yields:
\begin{align}
R_{\delta} &=\dfrac{S_{\textrm{ext}}^2 V}{2\pi} \left[\int^{\infty}_{0}\dfrac{\varrho d\varrho}{2\left(\varrho^2+1\right)^2}+O\left(V^2\right)\right]\ ,
\end{align}
which after integration is equal to:
\begin{equation}
R_{\delta}=\dfrac{S_{\textrm{ext}}^2 V}{8\pi}\left[1+O\left(V^2\right)\right]\ .
\label{2D:WRDirac0}
\end{equation}

Now, using the change of variables $w=\varrho\left(\varrho^2+1\right)/V$, and a series expansion of the integrand at $V\rightarrow\infty$, the integral in eq.~\eqref{2D:WRDirac} can be approximated by: 
\begin{align*}
\int^{\infty}_{0}\left[1-\dfrac{\varrho\left(\varrho^2+1\right)}{\sqrt{V^2+\varrho^2\left(\varrho^2+1\right)^2}}\right] \varrho^3 &=\dfrac{V^{4/3}}{3}\int^{\infty}_{0}w^{1/3}\left(1-\dfrac{w}{\sqrt{1+w^2}}\right)+O\left(V^{1/3}\right)\ , \\
&= \dfrac{\Gamma\left(\dfrac{1}{3}\right)\Gamma\left(\dfrac{7}{6}\right)}{4\sqrt{\pi}}V^{4/3}+O\left(V^{1/3}\right)\ ,
\end{align*}
which allows us to obtain eq.~\eqref{2D:WRDiracInf}.

\end{document}